\documentclass[twocolumn,showpacs,preprintnumbers,amsmath,amssymb,aps]{revtex4}
\usepackage{graphicx}% Include figure files
\usepackage{dcolumn}% Align table columns on decimal point
\usepackage{bm}% bold math

\newcommand{\lgun}{\lambda_{\rm gun}}
\newcommand{\Vgun}{V_{\rm gun}}
\newcommand{\Igun}{I_{\rm gun}}
\newcommand{\psigun}{\psi_{\rm gun}}

\newcommand{\Alfven}{Alfv\'{e}n\ }

\newcommand{\Bz}{B_{\rm Z}}
\newcommand{\Bphi}{B_\phi}

\newcommand{\Btor}{B_{\rm tor}}
\newcommand{\Bpol}{B_{\rm pol}}
\newcommand{\Bvec}{\mathbf{B}}
\newcommand{\Te}{T_{\rm e}}
\newcommand{\Ti}{T_{\rm i}}

\begin{document}

%\preprint{APS/123-QED}

\title{Experimental Identification of the Kink Instability as a
Poloidal Flux Amplification Mechanism for Coaxial Gun Spheromak Formation}

\author{S.~C.~Hsu}
\altaffiliation[Now at: ]{Los Alamos National Laboratory, Los Alamos,
NM\, 87545}
%\email{scotthsu@lanl.gov}
\author{P.~M.~Bellan}
%\email{pbellan@caltech.edu}
\affiliation{California Institute of Technology, Pasadena, CA\, 91125}
\date{April 28, 2003; accepted by Phys.\ Rev.\ Lett.}
%\date{\today}
%  It is always \today, today,

\begin{abstract}
The magnetohydrodynamic kink instability is observed and identified
experimentally as
a poloidal flux amplification mechanism for coaxial gun spheromak formation.
Plasmas in this experiment fall
into three distinct regimes which depend on the peak gun current to
magnetic flux ratio, with (I)~low values resulting in a straight
plasma column with helical magnetic field, (II)~intermediate values
leading to kinking of the column axis, and (III)~high values leading
immediately to a detached plasma.  Onset of column kinking agrees
quantitatively with the Kruskal-Shafranov limit, and the kink acts
as a dynamo which converts toroidal to poloidal flux.  
Regime~II clearly leads to both poloidal flux amplification
and the development of a spheromak configuration.
\end{abstract}


\maketitle

The spheromak~\cite{Jarboe94,Bellan00} is a simply-connected plasma
configuration in which the magnetic fields are largely determined by
dynamo-driven plasma currents.  Because of its topological simplicity
and ease of formation, the spheromak is of interest as a magnetic fusion
confinement scheme~\cite{Mclean02}.
Spheromak formation has traditionally been explained
by Taylor's hypothesis~\cite{Taylor86} that a turbulent
magnetohydrodynamic (MHD) system relaxes to a state of minimum
magnetic energy subject to the constraint of constant magnetic helicity.
While this hypothesis has successfully explained
the existence and many equilibrium properties of spheromaks,
it says nothing about the actual 3D dynamics underlying the
relaxation process, {\em e.g.}, for coaxial
gun spheromaks, the dynamo mechanism which converts injected
toroidal flux into required poloidal flux~\cite{al-Karkhy93}.  The
dynamics must be 3D because Cowling's theorem~\cite{Cowling34} shows
that purely axisymmetric processes cannot accomplish this.
This Letter experimentally identifies the MHD kink
instability~\cite{Freidberg87}
as a 3D mechanism which converts toroidal to poloidal flux
in a coaxial gun system, thereby leading to spheromak formation.
This mechanism should also be of fundamental importance to
coaxial helicity injection in spherical tori~\cite{Raman03},
relaxation in reversed-field pinches~\cite{Taylor86}, solar
coronal plasma instabilities~\cite{Vrsnak91}, and astrophysical
jets~\cite{Li01}.

Plasmas in this experiment fall into three regimes depending on peak $\lgun =
\mu_0 \Igun/\psigun$ (where $\Igun$ is the gun current and $\psigun$
is the bias poloidal magnetic flux intercepting the inner gun
electrode), with (I)~low values resulting in a straight plasma column
with helical magnetic field along the symmetry axis, (II)~intermediate
values leading to kinking of the column axis, and (III)~high values
leading immediately to a detached plasma with $\Btor \gg \Bpol$.
Onset of column kinking agrees quantitatively with the
Kruskal-Shafranov limit~\cite{Freidberg87}, and the kink
acts as a dynamo which converts toroidal to poloidal flux.
Regime~II clearly leads to
both poloidal flux amplification and magnetic field profiles
consistent with spheromak formation.  These results are qualitatively
consistent with recent time-dependent resistive MHD numerical
simulations of electrostatically driven spheromaks~\cite{Sovinec01}.

Early coaxial gun experiments~\cite{Lindberg64} demonstrated
poloidal flux amplification, and it was postulated that
this was due to an observed helical instability.
More recently, toroidal mode number $n=1$ central column
instabilities in coaxial gun
experiments were reported by several spheromak research
groups~\cite{Knox86,Browning92,Nagata93,Duck97,Brennan99}.  Typically,
the observations were based on edge magnetic measurements, and the
mode was studied in the context of relaxation during sustainment.  Two
hypotheses were proposed to explain the mode: (1)~development of
a $q=1$ surface in the closed-flux region~\cite{Knox86,Nagata93} or
equivalently a magnetic axis kink, and (2)~current-driven instability
of the central column~\cite{Duck97,Brennan99} or equivalently a
geometric axis kink.  Additionally, the mode was shown to couple
power from the central column to the spheromak~\cite{Browning92}.  The
present work offers a significant new result in that a nonlinear $n=1$
geometric axis kink is directly observed to produce the dynamo and
poloidal flux amplification leading to spheromak formation.

\begin{figure}
\includegraphics[width=2.7truein]{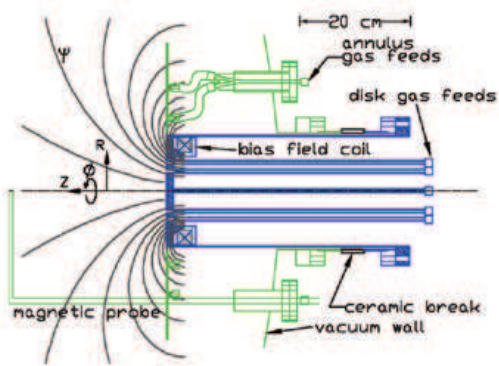}
\caption{\label{gun-probe} Side schematic of coaxial gun, showing bias
field coil and associated $\psi$-contours, gas feeds, and magnetic
probe.}
\end{figure}

Building on a prior Caltech spheromak formation
experiment~\cite{Yee00}, the present experiment simplifies, and makes
more accessible, the spheromak formation process by using a novel
coaxial gun~\cite{Hsu02a}.  A schematic of the gun is shown in
Fig.~\ref{gun-probe}, along with the cylindrical coordinate
system. The gun consists of two concentric, co-planar copper
electrodes:  a 20~cm diameter disk (blue) biased to negative polarity
high voltage surrounded by a 50~cm outer diameter annulus (green)
connected to vacuum chamber ground.  The width of the gap between
electrodes is 6~mm.  The inner electrode is mounted on the end of a
vacuum re-entrant port; both are electrically insulated from the
vacuum chamber by a ceramic break.  The co-planar arrangement of the
electrode surfaces is geometrically simple, allows the spheromak
formation process to be diagnosed with no mechanical obstructions, and
is feasible to model numerically~\cite{Tokman02}.   Bias field
poloidal flux $\psi$-contours are also shown in Fig.~\ref{gun-probe};
$\psigun$ can be adjusted from 0--7~mWb.  Both $\Igun$ and bias field
point toward the inner electrode.  The rise time of $\psigun$ is
10~ms, making it essentially stationary on the much faster time scale
of the gun discharge (tens of $\mu$s).  The gun is installed at one
end of a much larger vacuum chamber (about 2~m long and 1.5~m
diameter), and thus boundary effects on the spheromak formation
process are minimized. The gun is powered by an ignitron-switched
120~$\mu$F, 20~kV capacitor bank.  Hydrogen gas is injected
transiently using fast puff valves at eight equally spaced toroidal
positions on each electrode.  Due to the Paschen effect, the optimum
path for plasma breakdown is along the bias field and not at the gap
between electrodes.  The capacitor bank is discharged at $t=0$, at
which time the bias field and gas puff have already been introduced,
and breakdown occurs at approximately 4~$\mu$s.

\begin{figure}
\includegraphics[width=3.3truein]{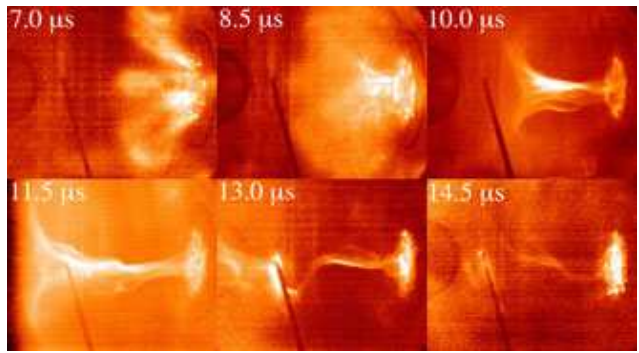}
\caption{\label{2472_images} Image sequence of shot 2472 taken with a
DRS Hadland Imacon 200 CCD camera (each frame originally 1200 by 980
pixels, 10 bits/pixel).  The circular gap between outer and inner
electrodes is visible toward right side of each frame.  $B$-probe is
also visible.  Kink is fully developed by 13~$\mu$s.}
\end{figure}

\begin{figure}
\includegraphics[width=3.3truein]{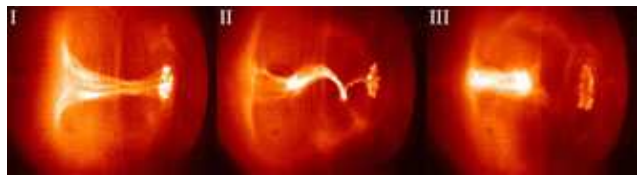}
\caption{\label{3_images} Images of three plasma regimes which depend
on increasing value of $\lgun$, taken with a Cooke Co.\ HSFC-PRO CCD
camera (each frame  originally 1280 by 1024 pixels, 12 bits/pixel):
(I)~stable column (shot 1210), (II)~kinked column (shot 1247), and
(III)~detached plasma (shot 1181).  $B$-probe was not installed for
these shots.}
\end{figure}

The main diagnostics are a multiple-frame fast CCD camera and a
60-channel magnetic probe array.  The camera takes a sequence of
time-resolved images in one plasma shot. The capability to visually
follow the global evolution of a plasma in one plasma shot is a new
and unique aspect of this work.  The inter-frame time is typically 1
or 1.5~$\mu$s, and the exposure time of each frame is 10 or 20~ns.
False color is applied to the images for viewing.  The magnetic probe
array, shown in Fig.~\ref{gun-probe}, measures all three components
($R$, $\phi$, $Z$) of $\Bvec$ at 20 radial positions with 2~cm radial
spacing.  Probe $\dot{B}$ signals are acquired using a digital
acquisition system and integrated numerically on a computer.  For all
$B$ measurements in this paper, the probe is located at $Z=22$~cm from
the plane of the gun electrodes.  Kink occurrence is independent of
the probe (see Figs.~\ref{2472_images} and \ref{3_images}).
Propagation of the plasma in $Z$ past the stationary probe relates
temporal information to $Z$-spatial information.  This relationship
improves as the propagation past the probe becomes fast compared to
the plasma expansion rate; this was exploited in the prior Caltech
spheromak experiment~\cite{Yee00}.  The time dependence thus acts as a
proxy for the $Z$-dependence.  $\Igun$ is measured with a Rogowski
coil surrounding the ceramic break, and gun voltage $\Vgun$ is
measured using a high-voltage probe.  Typical parameters are:  $\Vgun
= 4$--6~kV (charge voltage) and 2--2.5~kV  (after breakdown), peak
$\Igun = 70$--120~kA, $\psigun=0.5$--2~mWb, $B\approx 0.1$--1~kG,
$n\sim 10^{14}$~cm$^{-3}$, and $\Te\sim\Ti \approx 5-20$~eV.

Figure~\ref{2472_images} shows the time evolution of a kinked plasma
(shot 2472).  Gun electrodes are on the right side of each frame, and
the $Z$-axis is oriented horizontally across the middle of each frame.
At 7~$\mu$s, bright arches are visible soon after breakdown, showing
that breakdown occurs along the vacuum bias field lines.  As $\Igun$
increases, the arches expand and quickly merge together (8.5~$\mu$s),
forming a plasma column (10~$\mu$s) which begins to kink (11.5~$\mu$s)
and then becomes strongly kinked (13~$\mu$s).  It will be shown that
this sequence leads to a spheromak.  Depending on peak $\lgun$, three
distinct plasmas result, as shown in Fig.~\ref{3_images}.  Regime~I
leads to a stable column, II to a kinked column and then a spheromak,
and III to an immediately detached plasma.  The transition from
regime~I to II (II to III) occurs for peak $\lgun \approx 40$
(60)~m$^{-1}$.

\begin{figure}
\includegraphics[width=2.3truein]{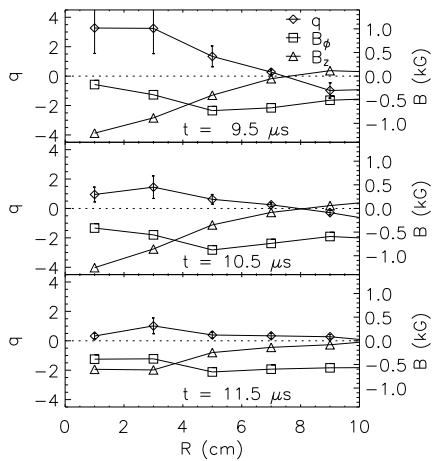}
\caption{\label{q-profiles} Flattening of $q(R)$ profile toward unity
(shot 2472) right before kink onset.  Error bars are due to
uncertainty in $L$. Also shown are $\Bphi$ and $\Bz$ profiles.}
\end{figure}

\begin{figure}
\includegraphics[width=2.1truein]{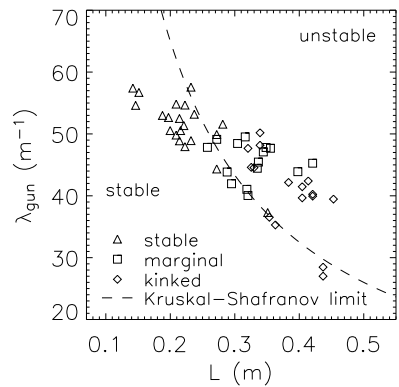}
\caption{\label{ksplot} Plot of $\lgun$ vs.\ column length $L$,
showing good quantitative agreement of kink onset with
Kruskal-Shafranov limit.}
\end{figure}

In order to show that the helical perturbation is an MHD kink mode,
consider the time evolution of the $q(R)$ profile (relative to
geometric axis), which is shown in Fig.~\ref{q-profiles} (shot 2472).
The kink mode becomes linearly unstable when $q(R)=2\pi R \Bz/L \Bphi
= 1$, where $L$ is the column length; this is the Kruskal-Shafranov
limit.  Since the kink involves a shift of the current channel, this
requires $q=1$ on axis and in its vicinity.  As seen in
Fig.~\ref{q-profiles}, $q$ near the axis is greater than unity at
9.5~$\mu$s but flattens and approaches unity by 11.5~$\mu$s,
coinciding with onset of column kinking (Fig.~\ref{2472_images}).  The
kink is fully developed at 13~$\mu$s and has broken apart by
14.5~$\mu$s.  The $q$ profile is determined using local $\Bz$ and
$\Bphi$ data from the magnetic probe array, and $L$ is determined from
the CCD images.  In addition, for an ensemble of plasma shots in which
a column forms (regimes I and II), the Kruskal-Shafranov limit
(recast~\cite{Hsu02b} as $\lgun = 4 \pi/L$) is a good
predictor~\cite{Hsu02b} of whether a kink actually develops
(Fig.~\ref{ksplot}).  Thus, two sets of independent data (magnetic
probe measurements of $\Bz$ and $\Bphi$, and Rogowski/flux loop
measurements of $\lgun$) both give direct evidence that the helical
perturbation is an MHD kink mode.  It is interesting to note that the
observed kinks always have one axial wavelength, even though there is
no fixed boundary at the end of the plasma column.  One
explanation~\cite{Nakamura01} for this is that there is a strong axial
gradient in $Bn^{-1/2}$, and thus an effective \Alfven speed
discontinuity, which acts as a rigid boundary.  This conjecture is
consistent with the ``mushroom cap'' on the left side of the kink in
Fig.~\ref{3_images}.

\begin{figure}
\includegraphics[width=2.1truein]{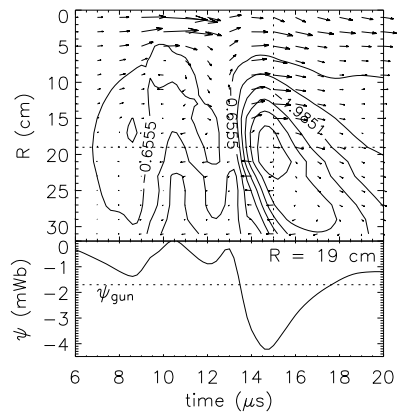}
\caption{\label{2472_vector} Plots (shot 2472) vs.\ $R$ and time of
(top)~$\Bpol$ vectors and $\psi$-contours (mWb), and (bottom)~$\psi$
along horizontal dashed line of top panel.}
\end{figure}

\begin{figure}
\includegraphics[width=2.1truein]{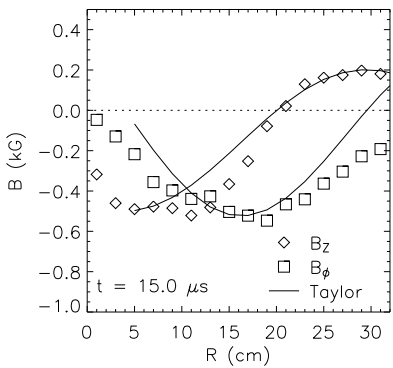}
\caption{\label{2472_b} Plots showing $B_{\rm Z}$, $B_\phi$ profiles
(shot 2472) compared to Taylor solutions ($\lambda = 15$~m$^{-1}$ and
radial offset of 4~cm).  Profiles are plotted for vertical dashed line
in Fig.~\ref{2472_vector} (top).}
\end{figure}

Next, it is shown that the kink is followed immediately by three
signatures of spheromak formation:  (1)~appearance of closed
$\psi$-contours [calculated assuming axisymmetry, $\psi(R,t) =
\int_0^R 2 \pi R' \Bz(R',t)\, {\rm d} R'$], (2)~$\psi$-amplification,
and (3)~magnetic field radial profiles  consistent with spheromak
formation.   It is important to note that the relationship between
closed $\psi$-contours and closed flux surfaces becomes ambiguous when
axisymmetry is broken.  Thus, in the presence of a non-axisymmetric
rotating kink, closed $\psi$-contours indicate closed flux surfaces
only in a time-averaged way.   Shortly after 13~$\mu$s, the kink
breaks apart (Fig.~\ref{2472_images}).  This coincides with signatures
of spheromak formation as observed in the magnetic probe measurements
(Fig.~\ref{2472_vector}).  At approximately 13~$\mu$s, closed
$\psi$-contours appear and $\psi_{\rm max}$ is amplified to larger
than $\psigun$ ($\psi$-amplification is due mainly to broadening of
the $\Bz$ profile after 12~$\mu$s).  At 15~$\mu$s, magnetic field
profiles consistent with spheromak formation are observed
(Fig.~\ref{2472_b}).  The measured radial profiles of $\Bz$ and
$\Bphi$ at 15~$\mu$s are compared with Taylor state
solutions~\cite{Taylor86} in cylindrical geometry, {\em i.e.}\ uniform
$\lambda$ solutions of
\begin{equation}
\nabla \times \Bvec = \lambda \Bvec.
\label{ff-eq}
\end{equation}
The solutions are $\Bz \sim J_0(\lambda R)$ and $\Bphi \sim
J_1(\lambda R)$, where $J_0$ and $J_1$ are Bessel functions of order
zero and one, respectively, and the best fit is found for $\lambda
\approx 15$~m$^{-1}$ with a radial offset of 4~cm (displacement of
spheromak off the geometric axis).  The slight disagreement between
measured profiles and the Taylor solution is not surprising since the
spheromak is expected to be either (1)~still undergoing relaxation
toward the Taylor state or (2)~in a modified relaxed state since it is
still being driven by the gun which has peak $\lgun \approx
50$~m$^{-1}$.

The kink modifies the direction of current-flow from $Z$
(poloidal) to $\phi$ (toroidal), as seen in Fig.~2. Equivalently, it
converts toroidal to poloidal flux. This process is
paramagnetic, amplifying $\psi$ over the initial applied
$\psigun=1.7$~mWb.  The paramagnetism is understood by realizing that
kinks involve perturbations with dependence ${\rm exp}(i\mathbf{k
\cdot x})$ where $\mathbf{k \cdot B}=0$.  The latter means $k_{\rm Z}
= -n\Bphi/R\Bz$. Thus, a locus of constant phase of the kink is given
by $\phi = (\Bphi /R \Bz)Z$, meaning that the kinked current channel
is a right (left) handed helix if $J_{\rm Z} \Bz > 0$ ($< 0$). This
will always lead to amplification of the original $\Bz$.  The
additional $\psi$ introduced by the kink can be estimated by
approximating the helix as a solenoid with current $I$, turns per
length $1/L$, and radius $a$.   The solenoid formula for the field
inside the solenoid is $\Bz=\mu_0 I/L$; the $\psi$ produced by the
solenoid is $\pi a^2 \Bz$ and thus depends non-linearly on the kink
amplitude $a$.  Using the measured values $a\approx 5$~cm, $L \approx
20$~cm, and $I\approx 60$~kA at 13.5~$\mu$s, the $\psi$ generated by
the kink is predicted to be approximately 1~mWb, which is within a
factor of 2 of the observed amplification of $\psi_{\rm max}$ over
$\psigun$.  The discrepancy is within the accuracy of $a$ and $L$
measurements and of the $\psi$ calculation assuming axisymmetry in the
presence of the rotating kink.  Because the coaxial gun
can only inject toroidal flux, 3D plasma dynamics must
be responsible for $\psi$ exceeding $\psigun$.  The dynamics are
provided by the kink, which amplifies
$\psigun$ by converting toroidal to poloidal flux; thus, in this
case, the kink constitutes the dynamo intrinsic to spheromak formation.

Regime~II clearly leads to all three signatures of spheromak
formation:  closed $\psi$-contours, $\psi$-amplification, and proper
magnetic field profiles.  Regime~I lacks both closed $\psi$-contours
and $\psi$-amplification.  Regime~III may have closed $\psi$-contours;
however, $\Btor \gg \Bpol$ associated with high $\Igun$, which could
indicate that the plasma is still relaxing. This work indicates
a close relationship between the kink threshold ($\lgun = 4\pi/L$) and
the spheromak formation threshold, which according to a {\em static}
force-free treatment is also proportional to an inverse length.  For
example, solutions of Eq.~(\ref{ff-eq}) for uniform $\lambda$ indicate
that closed $\psi$-contour equilibria exist for threshold $\lambda
\sim \sqrt{x_{11}^2/a^2 + \pi^2/h^2}$, where $a$ ($h$) is the radius
(length) of the flux conserver and $x_{11}$ is the first root of
$J_1$~\cite{Bellan00}.  Geometric differences among different
experiments could result in different eigenmode spectra of
Eq.~(\ref{ff-eq}) relative to the kink stability thresholds.

In summary, the MHD kink instability has been identified
experimentally as a poloidal flux amplification mechanism for coaxial
gun spheromak formation.  An $n=1$ central column helical instability
was observed during formation using multiple-frame CCD imaging. Onset
of the perturbation was shown using two independent sets of data to
agree quantitatively with the Kruskal-Shafranov limit.  The kink acts
as a dynamo which converts toroidal to poloidal flux, and
it is followed immediately by three key signatures of spheromak
formation: (1)~closed $\psi$-contours, (2)~$\psi$-amplification, and
(3)~magnetic field profiles similar to the Taylor state.

The authors acknowledge S.~Pracko, C.~Romero-Talam\'{a}s, and D.~Felt
for  technical assistance and H.~Bindslev for asking about kink
paramagnetism. Supported by a US-DoE Fusion Energy
Postdoctoral Fellowship and US-DoE Grant DE-FG03-98ER544561.

\end{document}